%% file: main.tex
\documentclass[10pt,twocolumn,letterpaper]{article}

\usepackage{iccv}
\usepackage{times}
\usepackage{epsfig}
\usepackage{graphicx}
\usepackage{float}
\usepackage{stfloats}
\usepackage{subfigure}
\usepackage{multirow}
\usepackage{amsmath}
\usepackage{amssymb}
\usepackage{cuted}
\usepackage{capt-of}
\usepackage[ruled]{algorithm2e}

\usepackage[breaklinks=true,bookmarks=false,colorlinks]{hyperref}

\iccvfinalcopy

\def\iccvPaperID{9135}

\ificcvfinal\fi

\begin{document}

\title{Image2Reverb: Cross-Modal Reverb Impulse Response Synthesis}

\author{Nikhil Singh\\
MIT\\
Media Lab\\
{\tt\small nsingh1@mit.edu}
\and
Jeff Mentch\\
Harvard University\\
SHBT\\
{\tt\small jsmentch@g.harvard.edu}
\and
Jerry Ng\\
MIT\\
Mechanical Engineering\\
{\tt\small jerryng@mit.edu}
\and
Matthew Beveridge\\
MIT\\
EECS\\
{\tt\small mattbev@mit.edu}
\and
Iddo Drori\\
MIT\\
EECS\\
{\tt\small idrori@mit.edu}
}

\maketitle
\ificcvfinal\thispagestyle{empty}\fi

\begin{strip}
    \centering
    \includegraphics[width=\textwidth]{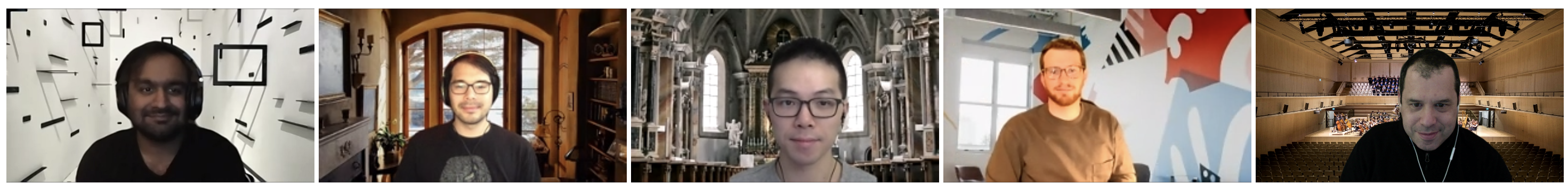}
    \captionof{figure}{An application of our work: matching speaker audio to expected reverberation of a video conference virtual background.\\ Project page with examples is available at \href{https://web.media.mit.edu/~nsingh1/image2reverb}{https://web.media.mit.edu/\raisebox{0.5ex}{\texttildelow}nsingh1/image2reverb}.}
    \label{fig:t}
\end{strip}

\ificcvfinal\thispagestyle{empty}\fi

\begin{figure*}[!ht]
    \centering
    \includegraphics[width=\textwidth]{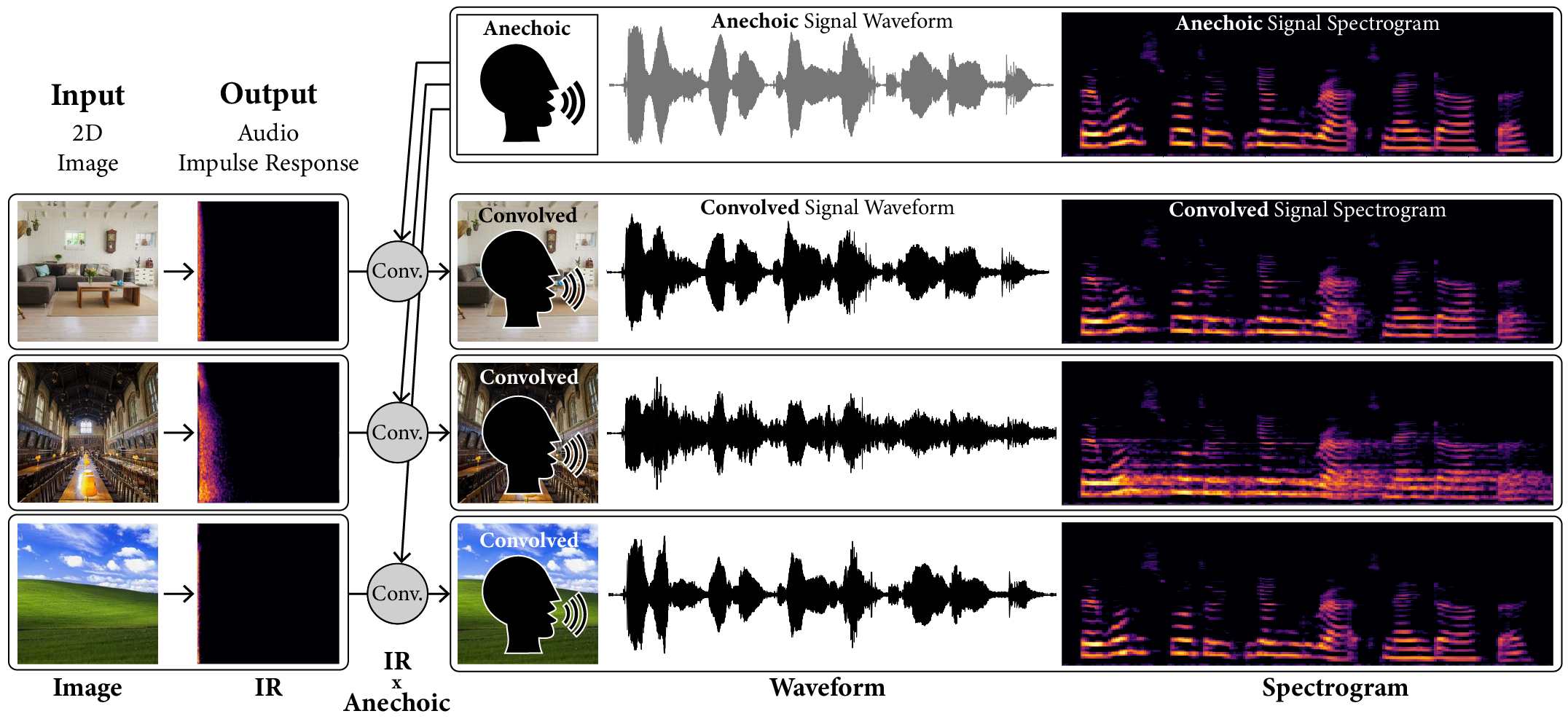}
    \caption{Generating audio impulse responses from images. Left: given an image of an acoustic environment as input, our model generates the corresponding audio impulse response as output. Right: generated impulse responses are convolved with an anechoic (free from echo) audio recording making that recording sound as if it were in the corresponding space. Waveforms and spectrograms are shown of the source anechoic signal and the same signal after convolution with the corresponding synthesized IR. All spectrograms are presented on a mel scale. Image2Reverb is the first system demonstrating end-to-end synthesis of realistic IRs from single images.}
    \label{fig:room2reverb_banner}
\end{figure*}

\begin{abstract}
Measuring the acoustic characteristics of a space is often done by capturing its impulse response (IR), a representation of how a full-range stimulus sound excites it. This work generates an IR from a single image, which can then be applied to other signals using convolution, simulating the reverberant characteristics of the space shown in the image. Recording these IRs is both time-intensive and expensive, and often infeasible for inaccessible locations. We use an end-to-end neural network architecture to generate plausible audio impulse responses from single images of acoustic environments. We evaluate our method both by comparisons to ground truth data and by human expert evaluation. We demonstrate our approach by generating plausible impulse responses from diverse settings and formats including well known places, musical halls, rooms in paintings, images from animations and computer games, synthetic environments generated from text, panoramic images, and video conference backgrounds.
\end{abstract}

\section{Introduction}
\label{introduction}
An effective and widely used method of simulating acoustic spaces relies on audio impulse responses (IRs) and convolution \cite{valimaki2012fifty, robjohns1999sony}. Audio IRs are recorded measurements of how an environment responds to an acoustic stimulus. IRs can be measured by recording a space during a burst of white noise like a clap, a balloon pop, or a sinusoid swept across the range of human hearing \cite{reilly1995convolution}. Accurately capturing these room impulse responses requires time, specialized equipment, knowledge, and planning. Directly recording these measurements may be entirely infeasible in continuously inhabited or inaccessible spaces of interest. End-to-end IR estimation has far ranging applications relevant to fields including music production, speech processing, and generating immersive extended reality environments. Our Image2Reverb system directly synthesizes IRs from images of acoustic environments. This approach removes the barriers to entry, namely cost and time, opening the door for a broad range of applications.

\begin{figure}[ht]
    \centering
    \includegraphics[width=0.4\textwidth]{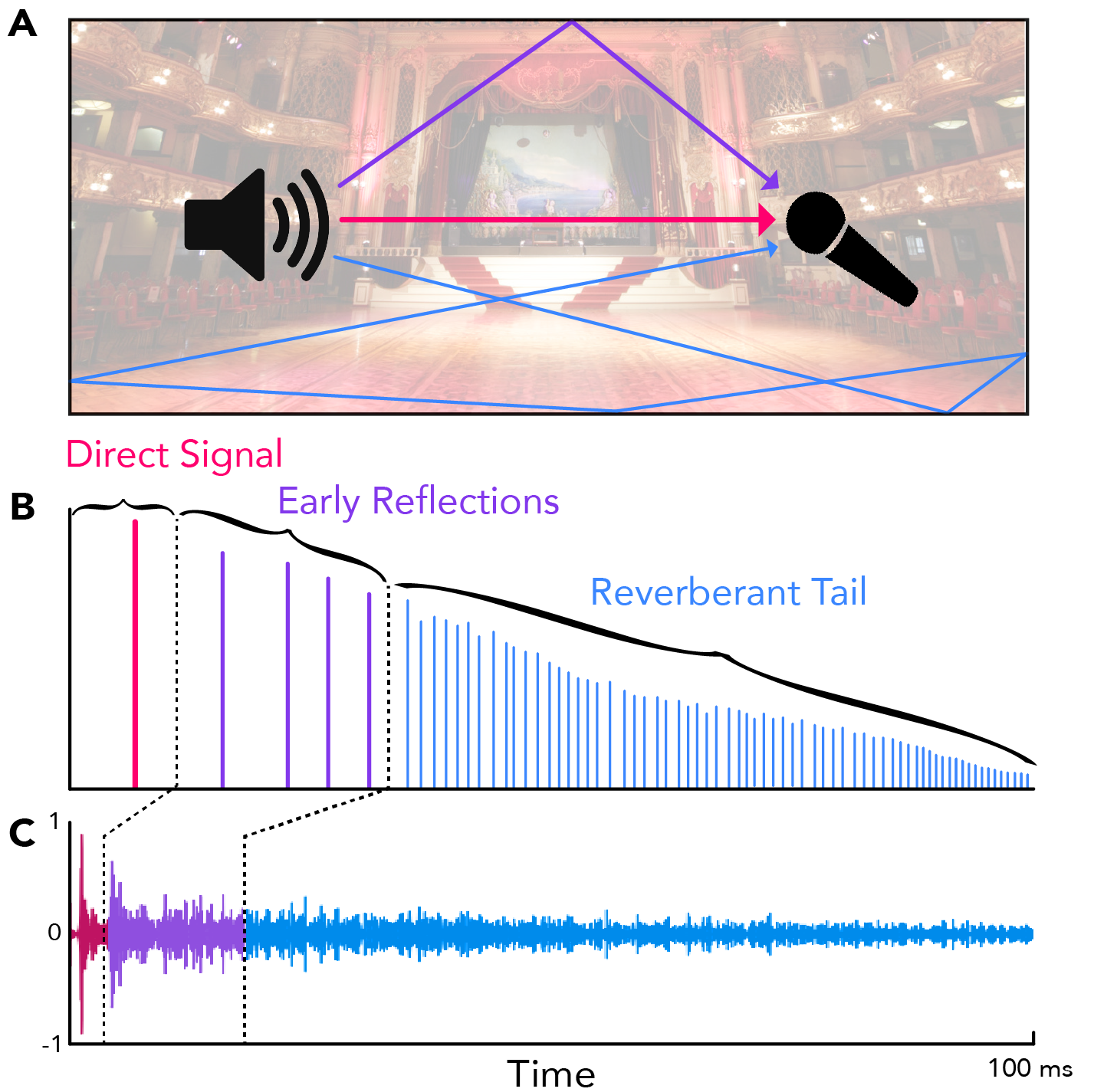}
    \caption{Impulse response overview. \textbf{(A)} Sound waves propagate across multiple paths as they interact with and reflect off their environment. These paths include the direct path from source to listener, early reflections including 1st and higher order reflections (after reflecting off 1 or more surfaces) and a more diffuse tail as they trail off and become more densely packed in time. These reflections make up the impulse response of the environment illustrated \textbf{(B)} schematically and \textbf{(C)} as a waveform.}
    \label{fig:ir_explanatory}
\end{figure}

In this work we model IR generation as a cross-modal paired-example domain adaptation problem and apply a conditional GAN \cite{goodfellow2014generative, gui2020review, mirza2014conditional} to synthesize plausible audio impulse responses conditioned on images of spaces. Next, we will describe related work that informs our approach.

\section{Related Work}
\label{previouswork}

\paragraph{Artificial reverberation.}
Historically, recording studios built reverberant chambers with speakers and microphones to apply reverb to pre-recorded audio directly within a physical space \cite{rettinger1957reverberation}. Reverberation circuits, first proposed in the 1960s, use a network of filters and delay lines to mimic a reverberant space \cite{Schroeder1961}. Later, Digital algorithmic approaches applied numerical methods to simulate similar effects. Conversely, convolution reverb relies on audio recordings of a space's response to a broadband stimulus, typically a noise burst or sine sweep. This results in a digital replica of a space's reverberant characteristics, which can then be applied to any audio signal \cite{anderegg2004convolution}.

Convolutional neural networks have been used for estimating late-reverberation statistics from images \cite{Kon2019, Kon2020}, though not to model the full audio impulse response from an image. This work is based on the finding that experienced acoustic engineers readily estimate a space's IR or reverberant characteristics from an image \cite{Kon2018}. Room geometry has also been estimated from 360-degree images of four specific rooms \cite{remaggi2019reproducing}, and used to create virtual acoustic environments which are compared with ground-truth recordings, though again IRs are not directly synthesized from the images. A related line of work synthesizes spatial audio based on visual information \cite{li2018, gao2019visual-sound, kim2019}. Prior work exists on synthesis of IRs using RNNs \cite{sali2020}, autoencoders \cite{steinmetz2018}, and GANs: IR-GAN \cite{ratnarajah2021} uses parameters from real world IRs to generate new synthetic IRs; whereas our work synthesizes an audio impulse response directly from an image.

\paragraph{Generative models for audio.}
Recent work has shown that GANs are amenable to audio generation and can result in more globally coherent outputs \cite{donahue2018adversarial}. GANSynth \cite{engel2018gansynth} generates an audio sequence in parallel via a progressive GAN architecture allowing faster than real-time synthesis and higher efficiency than the autoregressive WaveNet \cite{Oord2016} architecture. Unlike WaveNet which uses a time-distributed latent coding, GANSynth synthesizes an entire audio segment from a single latent vector. Given our need for global structure, we create a fixed-length representation of our input and adapt our generator model from this approach.

Measured IRs have been approximated with shaped noise \cite{lee2010approximating, bryan2020impulse}. While room IRs exhibit statistical regularities \cite{traer2016statistics} that can be modeled stochastically, the domain of this modeling is time and frequency limited \cite{badeau2019common}, and may not reflect all characteristics of real-world recorded IRs. Simulating reverb with ray tracing is possible but prohibitively expensive for typical applications \cite{Schissler2016}. By directly approximating measured audio IRs at the spectrogram level, our outputs are immediately applicable to tasks such as convolution reverb, which applies the reverberant characteristics of the IR to another audio signal.

\paragraph{Cross-modal translation.}
Between visual and auditory domains, conditional GANs have been used for translating between images and audio samples of people playing instruments \cite{Chen2017}. Our work builds on this by applying state-of-the-art architectural approaches for scene analysis and high quality audio synthesis, tuned for our purposes.

\section{Methods}
\label{methods}
\noindent Here we describe the dataset, model, and algorithm.

\subsection{Dataset}

\paragraph{Data aggregation.}
We curated a dataset of 265 different spaces totalling 1169 images and 738 IRs. From these, we produced a total of 11234 paired examples with a train-validation-test split of 9743-154-1957. These are assembled from sources including the OpenAIR dataset \cite{murphy2010openair}, other libraries available online, and web scraping. Many examples amount to weak supervision, due to the low availability of data: for example, we may have a ``kitchen" impulse response without an image of the kitchen in which it was recorded. In this case, we augmented with plausible kitchen scenes, judged by the researchers, gathered via web scraping and manual filtering. Although this dataset contains high variability in several reverberant parameters, e.g. early reflections and source-microphone distance, it allows us to learn characteristics of late-field reverberation.

\paragraph{Data preprocessing.}
Images needed to be filtered manually to remove duplicates, mismatches such as external pictures of an indoor space, examples with significant occlusive ``clutter" or excessive foreground activity, and intrusive watermarks. We then normalized, center-cropped at the max width or height possible, and downsampled to 224x224 pixels. We converted the audio IR files to monaural signals; in the case of Ambisonic B-Format sources we extracted the $W$ (omnidirectional) channel, and for stereo sources we computed the arithmetic mean of channels. In some cases, 360-degree images were available and in these instances we extract rectilinear projections, bringing them in line with the standard 2D images in our dataset.

\paragraph{Audio representation.}
Our audio representation is a log magnitude spectrogram. We first resampled the audio files to 22.050kHz and truncate them to 5.94s in duration. This is sufficient to capture general structure and estimate reverberant characteristics for most examples. We then apply a short-time Fourier transform with window size ($M=1024$) and hop size ($R=256$), before trimming the Nyquist bin, resulting in square 512x512 spectrograms. Finally, we take $\log(\lvert X \rvert)$ where $\lvert X \rvert$ represents the magnitude spectrogram; audio IRs typically contain uncorrelated phase, which does not offer structure we can replicate based on the magnitude.

\subsection{Model}

\paragraph{Components.}
Our model employs a conditional GAN with an image encoder that takes images as input and produces spectrograms. This overall design, with an encoder, generator, and conditional discriminator, is similar to that which Mentzer et al. \cite{mentzer2020high} applied to obtain state-of-the-art results on image compression, among many other applications. The generator and discriminator are deep convolutional networks based on the GANSynth \cite{engel2018gansynth} model (non-progressive variant), with modifications to suit our dataset, dimensions, and training procedure.

\begin{figure*}
    \centering
    \includegraphics[width=\textwidth]{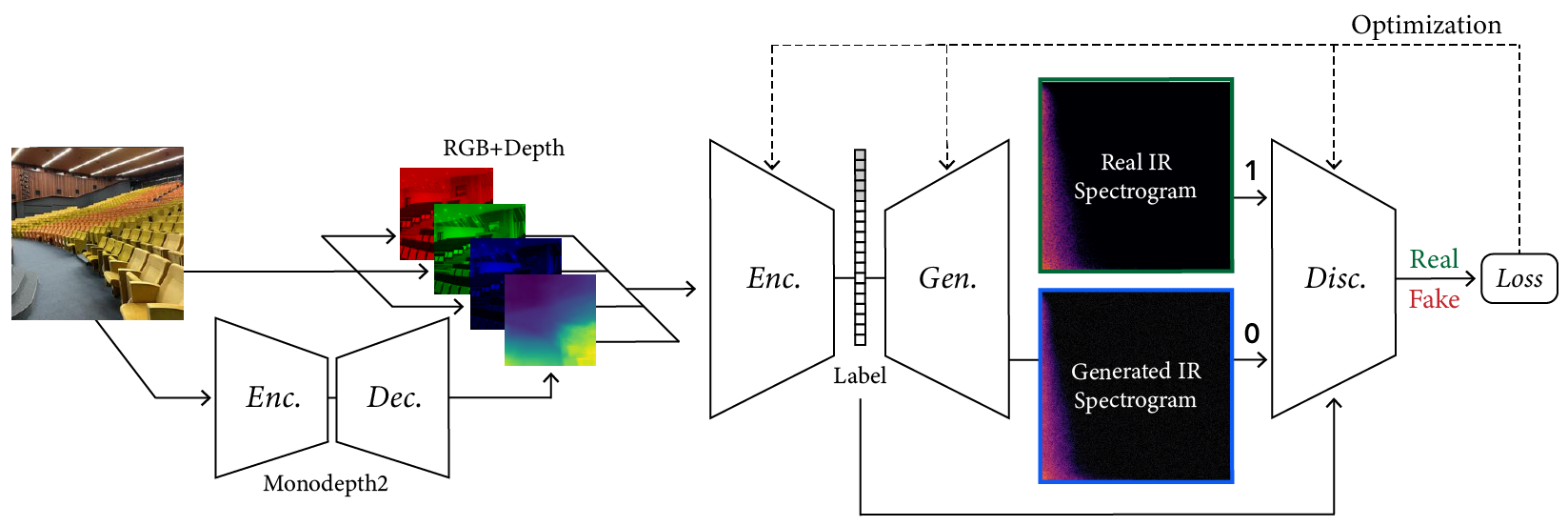}
    \caption{System architecture. Our system consists of autoencoder and GAN networks. Left: An input image is converted into 4 channels: red, green, blue and depth. The depth map is estimated by Monodepth2, a pre-trained encoder-decoder network. Right: Our model employs a conditional GAN. An image feature encoder is given the RGB and depth images and produces part of the Generator's latent vector which is then concatenated with noise. The Discriminator applies the image latent vector label at an intermediate stage via concatenation to make a conditional real/fake prediction, calculating loss and optimizing the Encoder, Generator, and Discriminator.}
    \label{fig:arch_summary}
\end{figure*}

The encoder module combines image feature extraction with depth estimation to produce latent vectors from two-dimensional images of scenes. For depth estimation, we use the pretrained Monodepth2 network \cite{monodepth2}, a monocular depth-estimation encoder-decoder network which produces a one-channel depth map corresponding to our input image. The main feature extractor is a ResNet50 \cite{he2016deep} pretrained on Places365 \cite{Zhou2014} which takes a four-channel representation of our scene including the depth channel (4x224x224). We add randomly initialized weights to accommodate the additional input channel for the depth map. Since we are fine-tuning the entire network, albeit at a low learning rate, we expect it will learn the relevant features during optimization. Our architecture's components are shown in Figure \ref{fig:arch_summary}.

\begin{algorithm}
    \caption{Forward and backward passes through the Image2Reverb model. Notation is explained in Table \ref{tab:vardefinitions}.}
    \label{algo:model}
    \KwIn{}
    Monodepth2: $x \sim X$; Encoder $\tilde{x} \sim \tilde{X}$; Generator: $z = E(\tilde{x}) \oplus u$; Discriminator: $(G(z), E(\tilde{x}))$ OR $(y, E(\tilde{x}))$;\\
    Parameters: (weight variables);\\
    \KwOut{}
    Monodepth2: $\mathbf{x_d}$; Encoder: $E(\tilde{x})$; Generator: G(z); Discriminator: $D(G(z), E(\tilde{x}))$ OR $D(y, E(\tilde{x}))$; \\
    \For{number of epochs}{
    Sample $B$ training images; \\
    Get depth $x_d = M(x)$; \\
    Append depth features to RGB channels ($y \oplus y_d$); \\
    Encoder image to feature-vector ($E(\tilde{x})$); \\
    Append noise ($z = E(\tilde{x}) \oplus u$;
    \\ Generate spectrogram ($G(z)$);
    \\ Forward pass through discriminator with either fake or real spectrogram ($D(G(z)| E(\tilde{x})$ OR $D(y | E(\tilde{x}))$);
    \\ Backward pass: update parameters for discriminator ($W_D$), generator ($W_G$), and encoder ($W_E$);
    }
\end{algorithm}

\begin{table}[ht]
    \small
    \centering
    \begin{tabular}{c|c}
        Notation & Definition \\
        \hline
         x & input image \\
         $x_d$ & estimated depth map \\
         $\oplus$ & concatenation operator\\
         $\tilde{x}$ & image with depth map ($x \oplus x_d$) \\
         $y$ & Real spectrogram \\
         $E, G, D$ & Encoder, Generator, Discriminator\\
         $M$ & Monodepth2 Encoder-Decoder \\
         $W_*$ & weights for a model \\
         $u$ & Noise, $u \sim \mathcal{N}(0, 1) $\\
         $z$ & Latent vector, encoder output and noise\\ & ($E(\tilde{x}) \oplus u$) \\
    \end{tabular}
    \caption{Notation and definitions for variables indicated in different parts of this paper.}
    \label{tab:vardefinitions}
\end{table}

\paragraph{Objectives.}
We use the least-squares GAN formulation (LSGAN) \cite{mao2017least}. For the discriminator:

\begin{align}
\begin{split}
    \underset{D}{\text{min}}\ V(D) &  = \mathbb{E}_{\mathbf{y} \sim p_{data}(\mathbf{y})}[(1 - D(y\ |\ E(\tilde{x}))^2]\\
    & + \mathbb{E}_{\mathbf{z} \sim p_{z}(\mathbf{z})}[(D(G(z)\ |\ E(\tilde{x}))^2]
\end{split}
\end{align}

For the generator, we introduce two additional terms to encourage realistic and high-quality output. First, we add an $\ell 1$ reconstruction term, scaled by a hyperparameter ($\lambda_a = 100$ in our case). This is a common approach in image and audio settings. Second, we introduce a domain-specific term that performs an estimation of the $T_{60}$ values, the time it takes for the reverberation to decay by $60dB$, for the real and generated samples, and returns the absolute percent error between the two scaled by a hyperparameter ($\lambda_b=100$ again). We term the differentiable $T_{60}$ proxy measure $T_{60_p}$. To compute this for log-spectrogram $x$, we first get the linear spectrogram $e^x$ and then sum along the time axis to obtain a fullband amplitude envelope. We use Schroeder's backward integration method to obtain a decay curve from the squared signal, and linearly extrapolate from the $-20dB$ point to get a $T_{60}$ estimate. In all:

\begin{align}
\begin{split}
    \underset{G}{\text{min}}\ V(G) &= \mathbb{E}_{\mathbf{z} \sim p_{z}(\mathbf{z})} \left[ \right. (1 - D(G(z)\ |\ E(\tilde{x}))^2\\ 
     & + \lambda_a \left\lVert G(z) - y \right\rVert_1 \\
     & + \left. \lambda_b \left\lvert \dfrac{T_{60_p}(G(z)) - T_{60_p}(y)}{T_{60_p}(y)} \right\rvert \right] 
\end{split}
\end{align}

\paragraph{Training.}
We train our model on 8 NVIDIA 1080 Ti GPUs. Three Adam optimizers for each of the Generator, Discriminator, and Encoder were used to optimize the networks' parameter weights. Hyperparameters are noted in Table \ref{tab:hyperparameters}. We make our model and code publicly available \footnote{Model and code: \href{https://github.com/nikhilsinghmus/image2reverb}{https://github.com/nikhilsinghmus/image2reverb}}.

\begin{table}[ht]
    \small
    \centering
    \begin{tabular}{c|c}
        Parameter & Value\\
        \hline
        $\eta_G$ & 4e-4\\
        $\eta_D$ & 2e-4\\
        $\eta_E$ & 1e-5\\
        $\beta$ & (0.0, 0.99)\\
        $\epsilon$ & 1e-8\\
    \end{tabular}
    \caption{Hyperparameters for the Generator, Discriminator, and Encoder initial learning rates, the optimizer beta ($\beta$), and epsilon ($\epsilon$) for the Adam optimizers we use (one each for $D, G, E$)}
    \label{tab:hyperparameters}
\end{table}

\section{Results}
Using Image2Reverb we are able to generate perceptually plausible impulse responses for a diverse set of environments. In this section, we provide input-output examples to demonstrate the capabilities and applications of our model and also review results of a multi-stage evaluation integrating domain-specific quantitative metrics and expert ratings. Our goal is to examine output quality and conditional consistency, generally considered important for conditional GANs \cite{devries2020on} and most relevant for our application.

\subsection{Examples}
\begin{figure}
    \centering
    \includegraphics[width=\columnwidth]{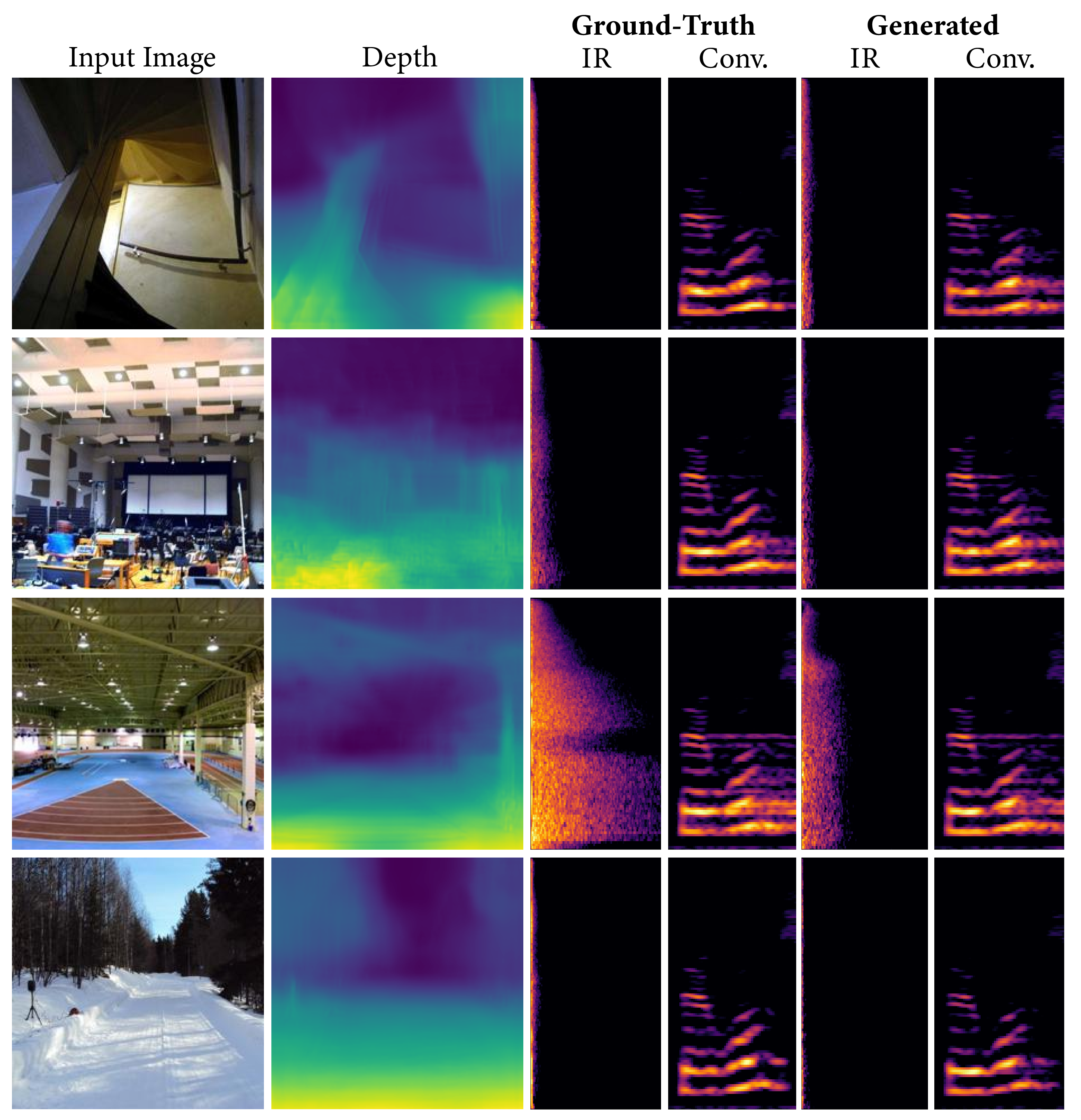}
    \caption{Ground-truth measured IRs vs generated IRs. Columns show input images, depth maps, measured IRs with corresponding convolved speech, and generated IRs with corresponding convolved speech. Larger indoor spaces here tend to exhibit greater $T_{60}$ times with longer measured impulse responses. The outdoor scene has a very short measured IR and corresponding generated IR. Input images are all examples that were used in the expert survey and were drawn from the test set.}
    \label{fig:p_groundtruth}
\end{figure}

\begin{figure}
    \centering
    \includegraphics[width=\columnwidth]{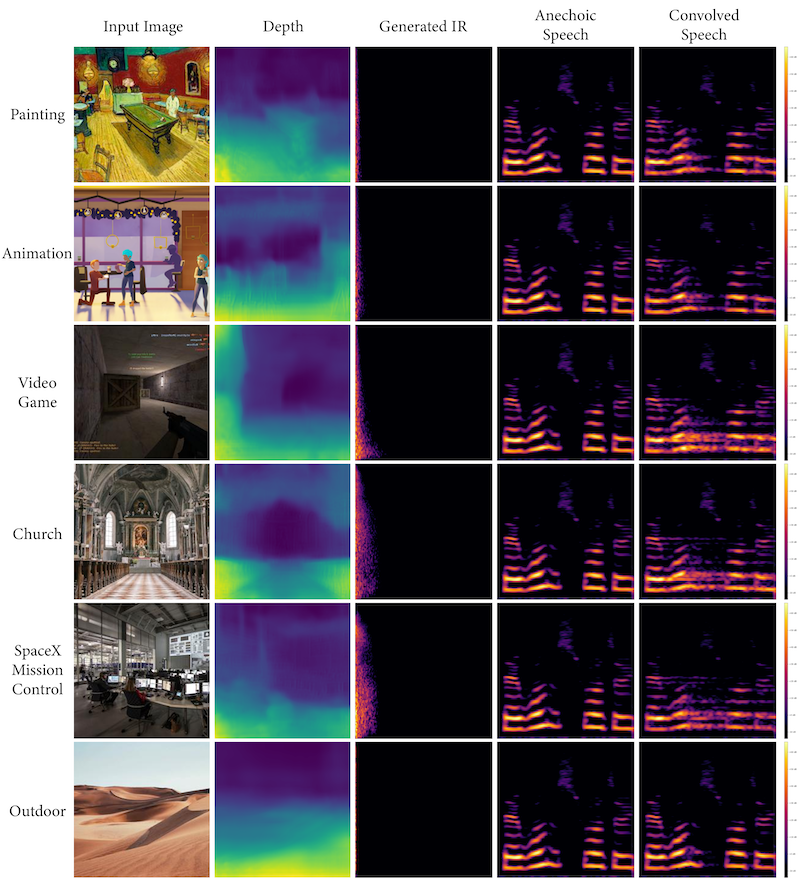}
    \caption{Generated IR examples. Columns show input images, depth maps, generated IRs, and a dry anechoic speech signal before and after the generated IR was applied via convolution. Input images come from a variety of spaces which illustrate possible applications of our model. Some images are synthetic, including: an oil painting, a 3D animation still, and a video game screenshot. Others come from real-world scenes like a church (where music is often heard), a famous yet inaccessible space (SpaceX), and an outdoor desert scene. Larger indoor spaces  tend to exhibit longer impulse responses as seen here.}
    \label{fig:p_all}
\end{figure}

\begin{figure}
    \centering
    \includegraphics[width=\columnwidth]{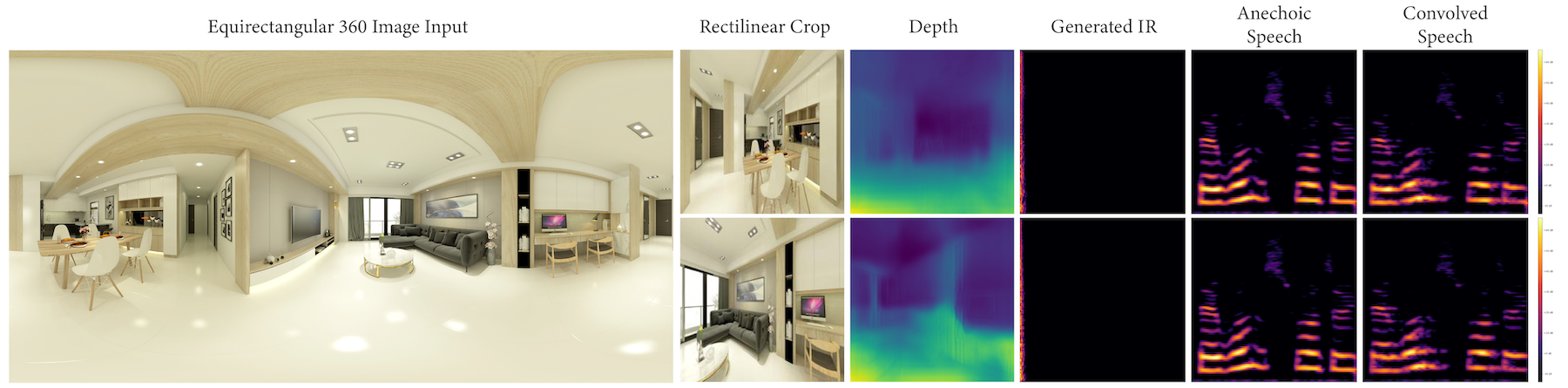}
    \caption{VR. Impulse responses generated from an equirectangular 360-degree image by sampling points on a sphere, cropping and applying a rectilinear projection to the resulting image, and feeding them into our model. This demonstrates how our model directly generates realistic impulse responses of panoramic virtual reality compatible images. Future work may allow generation of impulse responses using an entire 360-degree image, though at present there is a lack of paired data available for training.}
    \label{fig:p_vr}
\end{figure}

We present several collections consisting of diverse examples in our supplementary material, with inputs curated to illustrate a range of settings of interest including famous spaces, musical environments, and entirely virtual spaces. All examples are made available as audiovisual collections\footnote{Audiovisual samples: \href{https://web.media.mit.edu/~nsingh1/image2reverb/}{https://web.media.mit.edu/\raisebox{0.5ex}{\texttildelow}nsingh1/image2reverb/}} and were generated with a model trained in around 12 hours, with 200 epochs on a virtual machine. Figure \ref{fig:p_groundtruth} shows examples from our test set that were used in our expert evaluation (4 of 8, one from each category of: Small, Medium, Large, and Outdoor). We convolve a spoken word anechoic signal with the generated IRs for the reader to hear. Figure \ref{fig:p_all} takes images of diverse scenes (art, animation, historical/recognizable places) as inputs. Figure \ref{fig:p_vr} demonstrates how sections of 360-degree equirectangular images are cropped, projected, and passed through our model to generate IRs of spaces for immersive VR environments.

We strongly encourage the reader to explore these examples on the accompanying web page. We include examples of musical performance spaces, artistic depictions (drawings, paintings), 3D animation scenes, synthetic images from OpenAI's DALL•E, as well as real-world settings that present challenges (e.g. illusions painted on walls, reflections, etc.). These are largely created with real-world environments for which we may not have ground truth IRs, demonstrating how familiar and unusual scenes can be transformed in this way.

\subsection{Ablation Study}
\begin{figure}
    \centering
    \includegraphics[width=\columnwidth]{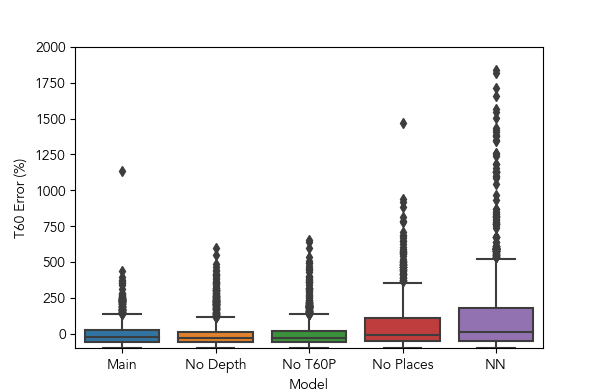}
    \caption{$T_{60}$ estimation error (\%) distributions from each model version. $T_{60}$ estimates how long it takes the reverberation to decay by $60dB$. ``Main" is our architecture as described earlier, ``No Depth" omits the depth maps, ``No T60P" omits the differentiable $T_{60_p}$ objective term, and ``No Places" uses randomly initialized encoder weights. ``NN" applies a nearest-neighbor approach with Places365-ResNet50 embeddings for images (errors clipped to 2000\% for clarity). Descriptive statistics are given in Table \ref{tab:modelcomparison_metrics}.}
    \label{fig:models_comparison}
\end{figure}

To understand the contribution of key architectural components and decisions, we perform a study to characterize how removing each affects test set $T_{60}$ estimation after 50 training epochs. The three components are the depth maps, the $T_{60_p}$ objective term, and the pretrained Places365 weights for the ResNet50 encoder. Figure \ref{fig:models_comparison} shows $T_{60}$ error distributions over the test set for each of these model variants, and Table \ref{tab:modelcomparison_metrics} reports descriptive statistics.

Our model reflects better mean error (closer to 0\%) and less dispersion (a lower standard deviation) than the other variants. The former is well within the just noticeable difference (JND) bounds for $T_{60}$, often estimated as being around 25-30\% for a musical signal \cite{jndt60}. Additionally, this is an upper bound on authenticity: a more rigorous goal then perceptual plausibility \cite{pellegrini2001quality}. The lower standard deviation indicates generally more consistent performance from this model across different examples, even in the presence of some that cause relatively large estimation errors due to incorrect interpretation of relevant qualities in the image, or inaccurate/noisy synthesis or estimation.

\begin{table}
    \small
    \centering
    \begin{tabular}{ccccccc}
        & & Main & -Depth & -$T_{60_p}$ & -P365 & NN\\
        \hline
        \multirow{4}{*}{\begin{tabular}{c} $T_{60}$ Err\\ (\%)\\ \  \end{tabular}}
        & $\mu$ & \textbf{-6.03} & -9.17 & -7.1 & 43.15 & 149\\
        & $\sigma$ & \textbf{78.8} & 83.1 & 85.97 & 144.3 & 491.02\\\\
    \end{tabular}
    \caption{$T_{60}$ estimation error (\%) statistics from each model version. ``Main" is our architecture as described earlier, ``-Depth" omits depth maps,``-$T_{60_p}$" omits the differentiable $T_{60_p}$ objective term, and ``-P365" does not use the pretrained Places365 weights for the ResNet50 encoder. ``NN" indicates a nearest-neighbor approach with Places365-ResNet50 embeddings for images. For mean and median, values closer to 0 reflect better performance. For the standard deviation, lower values reflect better performance. Distributions are visualized in Figure \ref{fig:models_comparison}.}
    \label{tab:modelcomparison_metrics}
\end{table}

\subsection{Expert Evaluation}
Following the finding that experienced acoustic engineers readily estimate a space's reverberant characteristics from an image \cite{Kon2018}, we designed an experiment to evaluate our results. We note that this experiment is designed to estimate comparative perceptual plausibility, rather than (physical) authenticity (e.g. by side-by-side comparison to assess whether any difference can be heard). These goals have been differentiated in prior work \cite{pellegrini2001quality}. We selected two arbitrary examples from each of the four scene categories and recruited a panel of 31 experts, defined as those with significant audio experience, to participate in a within-subjects study. For each of these examples, we convolved an arbitrary anechoic signal with the output IR, as well as the ground truth IR. These 16 samples were presented in randomized order and participants were instructed to rate each on a scale from 1 to 5 based on 1) reverberation quality, and 2) realism or ``match" between their expected reverb based on the image and the presented signal with reverb applied. Participants answered one reverb-related screening question to demonstrate eligibility, and two attention check questions at the end of the survey. The four scene categories are: Large, Medium, Outdoor, and Small. These demonstrate diversity in visual-reverb relationships. The dependent variables are quality and match ratings, and the independent variables are IR source (real vs. fake) and scene category (the four options listed previously). We first test our data for normality with D'Agostino and Pearson's omnibus test \cite{pearson1977tests}, which indicates that our data is statistically normal ($p>.05$).

\begin{figure}
    \centering
    \includegraphics[width=\columnwidth]{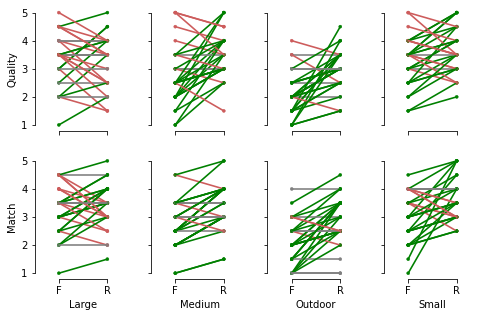}
    \caption{Expert evaluation results. Paired plots showing per-participant quality and match differences in rating for each scene category. Green lines indicate higher rating for real IRs, red lines for generated IRs, and grey lines equivalent ratings for both.}
    \label{fig:ratings}
\end{figure}

A two-way repeated-measures ANOVA revealed a statistically significant interaction between IR source and scene category for both quality ratings, $F(3, 90)=7.04$, $p\leq .001$, and match ratings, $F(3, 90)=3.73$, $p=.02$ (reported $p$-values are adjusted with the Greenhouse-Geisser correction \cite{greenhouse1959methods}). This indicates that statistically significant differences between ratings for real and fake IR reverbs depend on the scene category. Per-participant ratings and rating changes, overall and by scene, are shown in Figure \ref{fig:ratings}.

Subsequent tests for simple main effects with paired two one-sided tests indicate that real vs. fake ratings are statistically equivalent ($p<.05$) for large and small quality ratings, and large, medium, and small match ratings. These tests are carried out with an $\epsilon$ of 1 (testing for whether the means of the two populations differ by at least 1). Results are shown in Table \ref{tab:ratings_maineffects}. Notably, outdoor scenes appear to contribute to the rating differences between real and fake IRs. We conjecture this is due to outdoor scenes being too different a regime from the vast majority of our data, which are indoor, to model effectively. Additionally, medium-sized scenes appear to contribute to differences in quality.

\begin{table}
    \small
    \centering
    \begin{tabular}{llcc}
        Rating & Scene & DoF & $p$\\
        \hline
        Quality & Large & 56 & \textbf{$<.001$} \\
        Quality & Medium & 56 & $.28$ \\
        Quality & Outdoor & 56 & $.62$ \\
        Quality & Small & 56 & \textbf{$<.001$} \\
        Match & Large & 56 & \textbf{$<.001$} \\
        Match & Medium & 56 & \textbf{$.006$} \\
        Match & Outdoor & 56 & $.29$ \\
        Match & Small & 56 & \textbf{$<.05$} \\
        
    \end{tabular}
    \caption{Simple main effect tests for equivalence between real and generated IRs across different categories of scenes. We use paired two one-sided tests with bounds ($\epsilon$) of 1 and Bonferroni-adjusted p-values. These results suggest that real vs. fake ratings are statistically equivalent within one rating unit (the resolution of the rating scale) for large and small quality ratings, and large, medium, and small match ratings. Notably, outdoor scenes contribute to the difference between real and fake IRs and medium-sized scenes contribute to differences in quality.}
    \label{tab:ratings_maineffects}
\end{table}

\subsection{Model Behavior and Interpretation}

\begin{figure}
    \centering
    \includegraphics[width=0.4\textwidth]{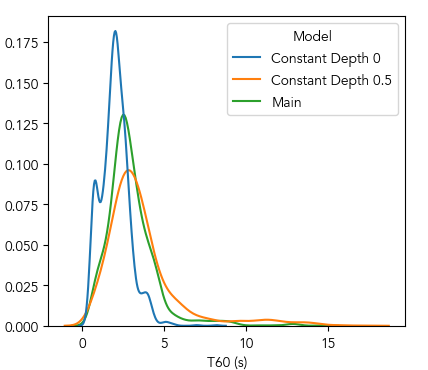}
    \caption{Effect of Depth on $T_{60}$. Distributions of estimated $T_{60}$ values for the model with estimated depth maps, plus constant depth maps set to either 0 (low) or 0.5 (high). Manipulating the depth value allows us to ``suggest" smaller or larger scenes, i.e. bias the output of the model. Table \ref{tab:t60_distributions} shows corresponding descriptive statistics. These results indicate a level of ``steerability" for the model's behavior in human-in-the-loop settings.}
    \label{fig:t60_distributions}
\end{figure}

\paragraph{Effect of varying depth.}
We compare the full estimated depth map with constant depth maps filled with either 0 or 0.5 (chosen based on the approximate lower and upper bounds of our data). We survey the distributions of generated IRs' $T_{60}$ values over our test set, the results of which are shown in Figure \ref{fig:t60_distributions}. Table \ref{tab:t60_distributions} reports descriptive statistics for these distributions, showing that the main model's output IRs' decay times are biased lower by the 0-depth input and higher by the 0.5-depth input respectively. These may indicate some potential for steering the model in interactive settings. We do note, however, that behavior with constant depth values greater than 0.5 is less predictable. This may be due to the presence of outdoor scenes, for which the scene's depth may not be correlated with IR duration. 

\begin{table}
    \small
    \centering
    \begin{tabular}{ccccc}
        & & Main & Depth 0 & Depth 0.5\\
        \hline
        \multirow{4}{*}{$T_{60}$ (s)}
        & $\mu$ & 2.07 & 2.01 & 3.62\\
        & $\sigma$ & 1.54 & 0.87 & 2.36\\
        & $Mdn.$ & 2.69 & 2.00 & 3.07\\
    \end{tabular}
    \caption{Descriptive statistics for the model with estimated depth maps, as well as constant depth maps set to either 0 or 0.5. The full depth map's results are between that of the 0 and 0.5 depth maps. Figure \ref{fig:t60_distributions} visulizes the corresponding distributions.}
    \label{tab:t60_distributions}
\end{table}

\paragraph{Effect of transfer learning.}
To understand which visual features are important to our encoder, we use Gradient-weighted Class Activation Mapping (Grad-CAM) \cite{selvaraju2017grad}. Grad-CAM is a popularly applied strategy for visually interpreting convolutional neural networks by localizing important regions contributing to a given target feature (or class in a classification setting). We produce such maps for our test images with both the ResNet50 pre-trained on Places365 dataset, as well as the final encoder model. All resulting pairs exhibit noticeable differences; we check for this with the structural similarity index (SSIM) metric \cite{wang2004image}, which is below 0.98 for all examples.

\begin{figure}[ht]
    \centering
    \includegraphics[width=0.4\textwidth]{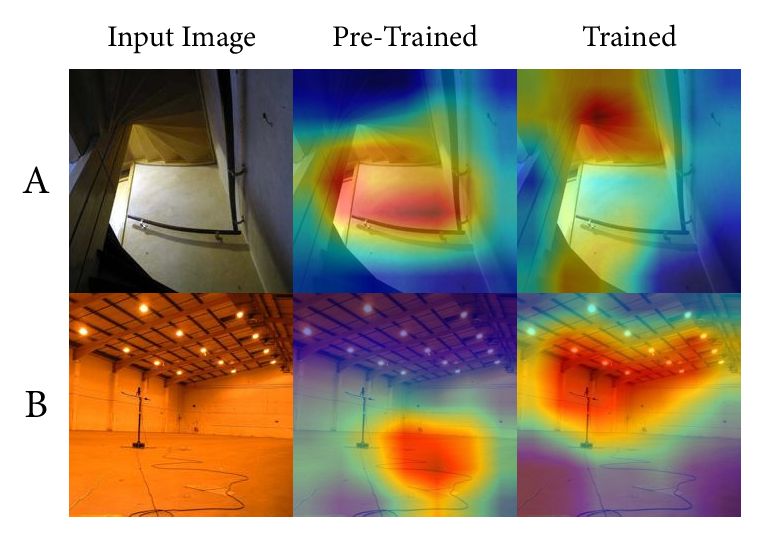}
    \caption{Grad-CAMs for images passed through the pre-trained Places365 ResNet50 encoder vs. our fine-tuned encoder, showing movement towards significant reflective areas for \textbf{(A)} a small, and \textbf{(B)} a large environment. The fine-tuned model's activations highlight larger reflective surfaces: depth of staircase for \textbf{(A)} vs. railing that may be more optimal for scene identification, and wall-to-ceiling corner plus surrounding areas for (B).}
    \label{fig:gradcam_reflective}
\end{figure}

\begin{figure}[ht]
    \centering
    \includegraphics[width=0.4\textwidth]{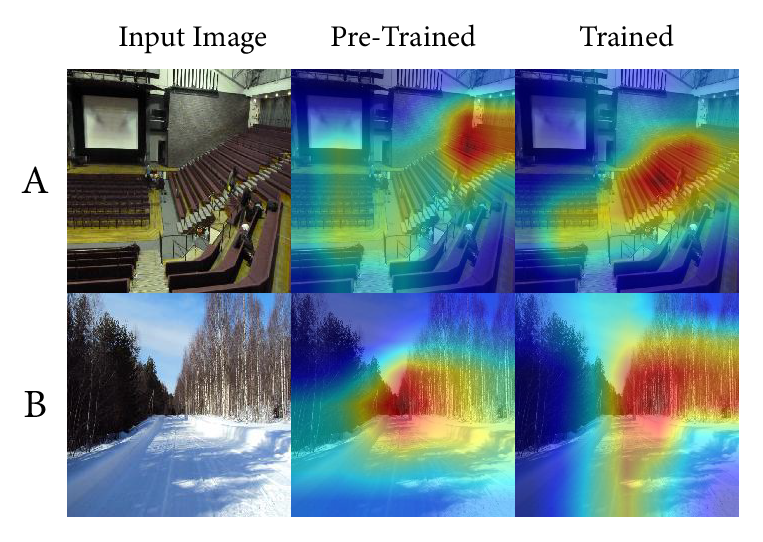}
    \caption{Grad-CAMs for images passed through both the pre-trained Places365 ResNet50 encoder and our fine-tuned encoder, showing movement towards more textured areas for \textbf{(A)} an indoor, and \textbf{(B)} an outdoor environment. The former seems to contain significant absorption and the latte has few reflective surfaces. In both cases, textured areas are highlighted. These may be associated with absorption, diffusion, and more sparse reflections depending on the scene.}
    \label{fig:gradcam_textured}
\end{figure}

We qualitatively survey these and identify two broad change regimes, which are illustrated with particular examples. First, we observe that the greatest-valued feature is often associated with activations of visual regions corresponding to large reflective surfaces. Examples are shown in Figure \ref{fig:gradcam_reflective}. Often, these are walls ceilings, windows, and other surfaces in reflective environments. Second, we find that textured areas are highlighted in less reflective environments. Examples of these are shown in Figure \ref{fig:gradcam_textured}. These may correspond to sparser reflections and diffusion.

\paragraph{Limitations and future work.}
Many images of spaces may offer inaccurate portrayals of the relevant properties (size, shape, materials, etc.), or may be misleading (examples in supplementary material), leading to erroneous estimations. Our dataset also contains much variation in other relevant parameters (e.g. $DRR$ and $EDT$) in a way we cannot semantically connect to paired images, given the sources of our data. New audio IR datasets collected with strongly corresponding photos may allow us to effectively model these characteristics precisely.

\section{Conclusion}
We introduced Image2Reverb, a system that is able to directly synthesize audio impulse responses from single images. These are directly applied in downstream convolution reverb settings to simulate depicted environments, with applications to XR, music production, television and film post-production, video games, videoconferencing, and other media. Our quantitative and human-expert evaluation shows significant strengths, and we discuss the method's limitations. We demonstrate that end-to-end image-based synthesis of plausible audio impulse responses is feasible, given such diverse applications. We hope our results provide a helpful benchmark for the community and future work and inspire creative applications.

{\small
\paragraph{Acknowledgements}
We thank the reviewers for their thorough feedback and useful suggestions. Additionally, we thank James Traer, Dor Verbin, and Phillip Isola for helpful discussions. We thank Google for a cloud platform education grant.
}

{\small
\bibliographystyle{ieee_fullname}
\bibliography{main}
}

\appendix
\input{supplementary}

\end{document}

%% file: supplementary.tex
\section{Supplementary Material}

\begin{figure*}[!hb]
    \centering
    \includegraphics[width=\textwidth]{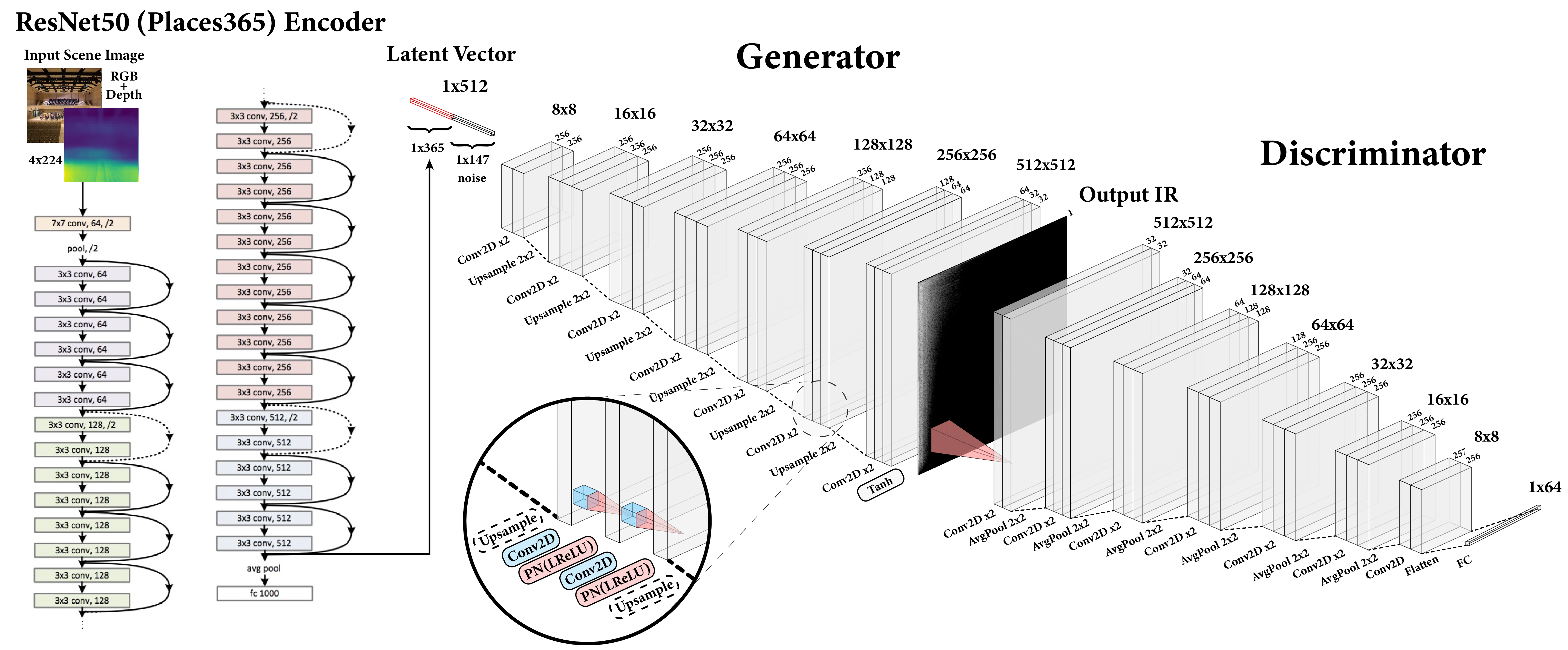}
    \caption{Detailed overview of Image2Reverb model architecture. Left: the ResNet50 encoder pre-trained on Places365 (figure at left adapted from \cite{he2016deep}). Right: the generator and discriminator. The output of the encoder consists of 365 features, to which we concatenate noise to produce a 512d latent vector. The generator and discriminator contain upsampling and downsampling convolutions respectively. A leaky rectified linear unit (LReLU), with $\alpha=0.2$, is used after each convolutional layer in the model in both the discriminator and the generator with the final layer of the generator using a $\tanh$ activation. PN denotes pixelwise normalization, which we use in the generator. The composition of blocks is based on ProGAN \cite{karras2018progressive}. The final step in the discriminator is a fully connected layer with a linear activation (scalar output).}
    \label{fig:nn_arch}
\end{figure*}

As supplementary material, we present and review a number of input/output examples across several categories with distinct properties\footnote{Link to audiovisual examples page: \href{https://web.media.mit.edu/~nsingh1/image2reverb/}{https://web.media.mit.edu/~nsingh1/image2reverb/}}. A summary of these results is shown in Table \ref{tab:results}. We additionally present a more detailed diagram of our architecture, shown in Fig. \ref{fig:nn_arch}.

\begin{figure*}
    \centering
    \includegraphics[width=\textwidth]{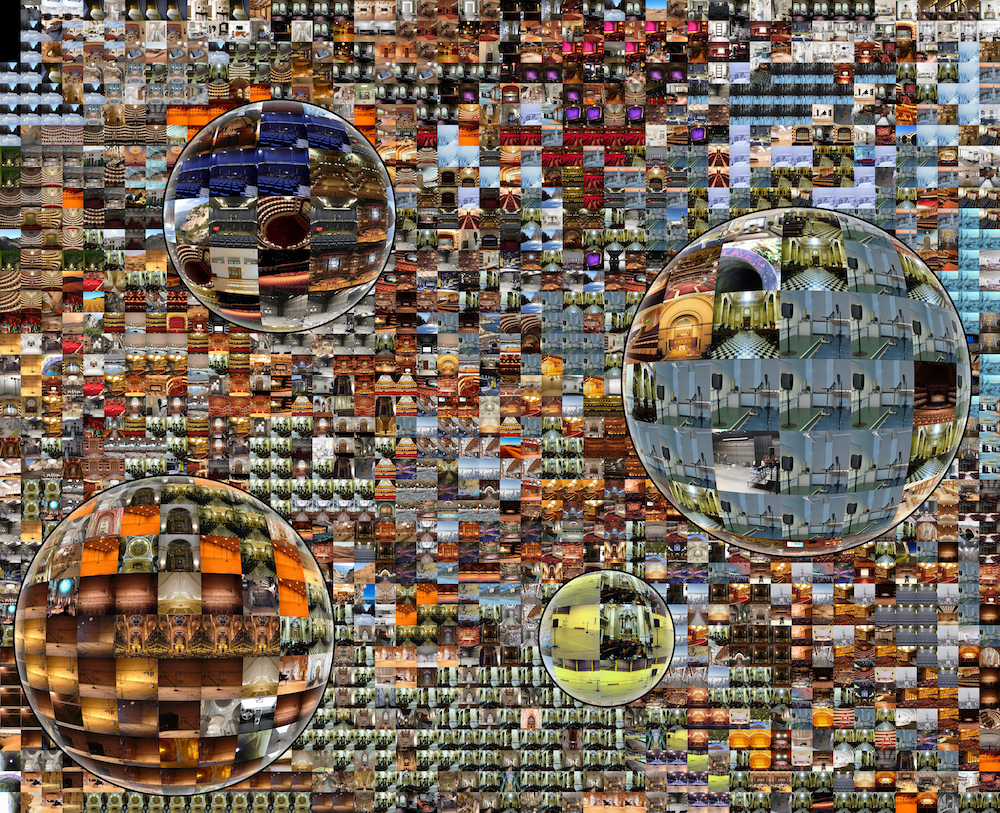}
    \caption{Manifold-based visualization of our test set. We compute multi-band $T_{60}$ estimates for output audio IRs for each image, and then perform nonlinear dimensionality reduction with t-SNE to obtain two-dimensional feature vectors for each example. We produce a grid by solving a linear assignment problem, as is commonly done to visualize large image datasets. Our visualization shows local clusters of same and similar scenes in many cases, but also some variation within scenes. In some outdoor settings, this variation grows considerably large, resulting in increased scattering. In other cases, we observe closeness between different views of the same scene and similar scenes.}
    \label{fig:tsne}
\end{figure*}

Finally, to gain a qualitative view of intra-scene and adjacent-scene consistency, we plot our test set input images according to the corresponding output audio characteristics by a visualization shown in Figure \ref{fig:tsne}. We produce multiband $T_{60}$ estimations from all output IRs, and then used t-SNE \cite{maaten2008visualizing} to reduce the data dimensionality to two dimensions. We then solve a linear assignment problem to transform this into a grid representation. Several instances of within-scene clusters are visible, as well as closeness of related scenes. This suggests that while our method does make errors (outliers are also visible), it learns to treat similar scenes similarly while capturing variation.

\begin{table}[!h]
    \centering
    \begin{tabular}{l|c|c}
         \textbf{Topic} &  \textbf{Figure \#} & \textbf{Images}\\
         \hline
         Famous and iconic places & \ref{fig:p_famous} & 6\\
         Musical environments & \ref{fig:p_music} & 6\\
         Artistic renderings & \ref{fig:p_art} & 6\\
         DALL•E-generated spaces & \ref{fig:p_dall-e} & 6\\
         Limitations (i.e. challenging examples) & \ref{fig:p_reflection} & 4\\
         Animated scenes & \ref{fig:p_animation} & 6\\
         Virtual backgrounds & \ref{fig:p_zoom} & 6\\
         Historical places & \ref{fig:p_other} & 5\\
         Video games & \ref{fig:p_videogame} & 4\\
         Common and identifiable scenes & \ref{fig:p_everyday} & 6\\
         \hline
         Total & & 55\\
    \end{tabular}
    \caption{Additional Results.}
    \label{tab:results}
\end{table}

\begin{figure*}[!hb]
    \centering
    \includegraphics[width=0.9\textwidth]{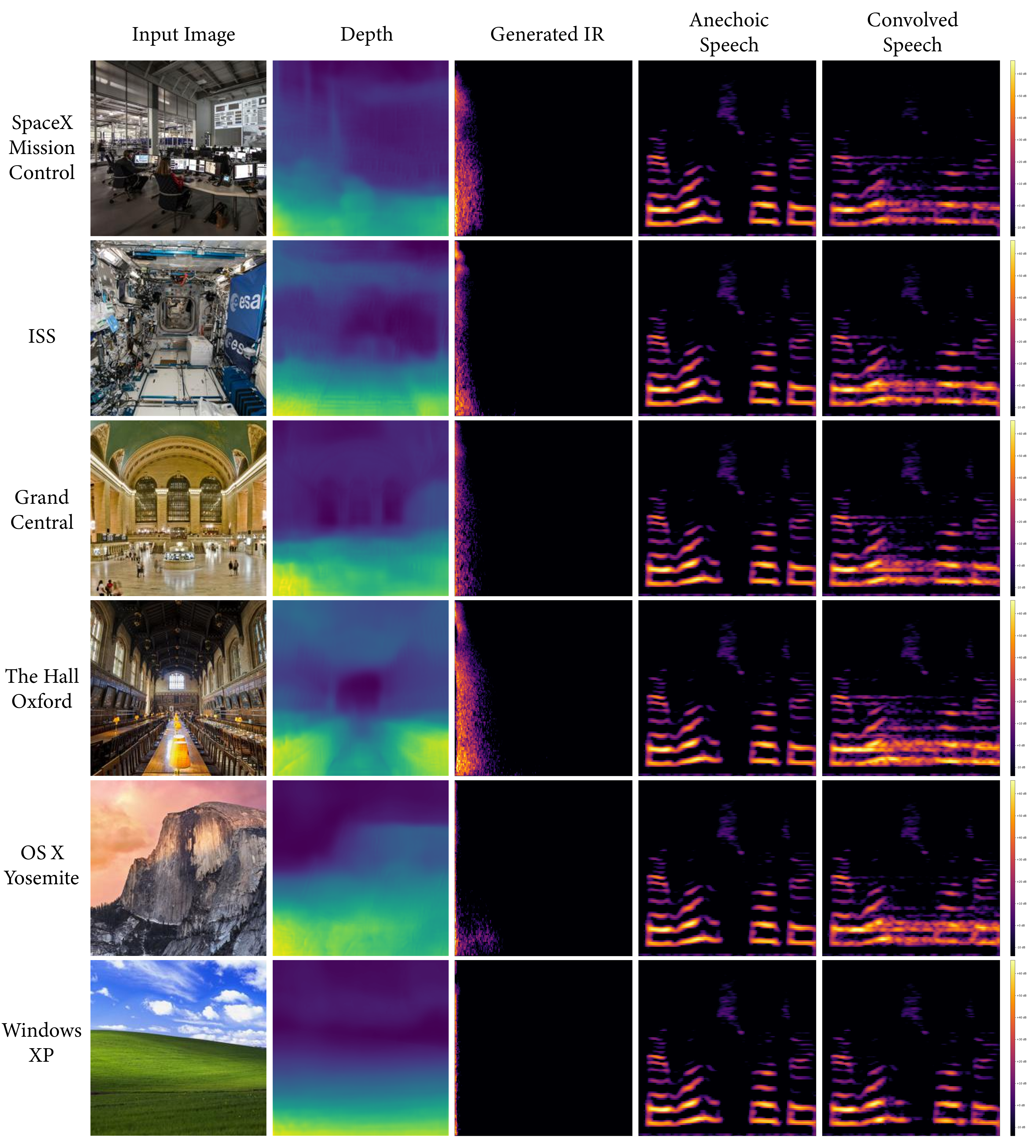}
    \caption{Famous and iconic spaces. Columns show input images, depth maps, generated IRs, and a dry anechoic speech signal before and after the generated IR was applied to the signal via convolution. The input images come from spaces that may be impractical or impossible to record in. The indoor spaces here show longer impulse responses compared to the outdoor scenes which is typically observed and expected in real-world settings. Larger indoor spaces also tend to exhibit greater $T_{60}$ times with longer impulse responses which we see here, though the ISS image has a longer impulse response than we expect.}
    \label{fig:p_famous}
\end{figure*}

\begin{figure*}
    \centering
    \includegraphics[width=0.9\textwidth]{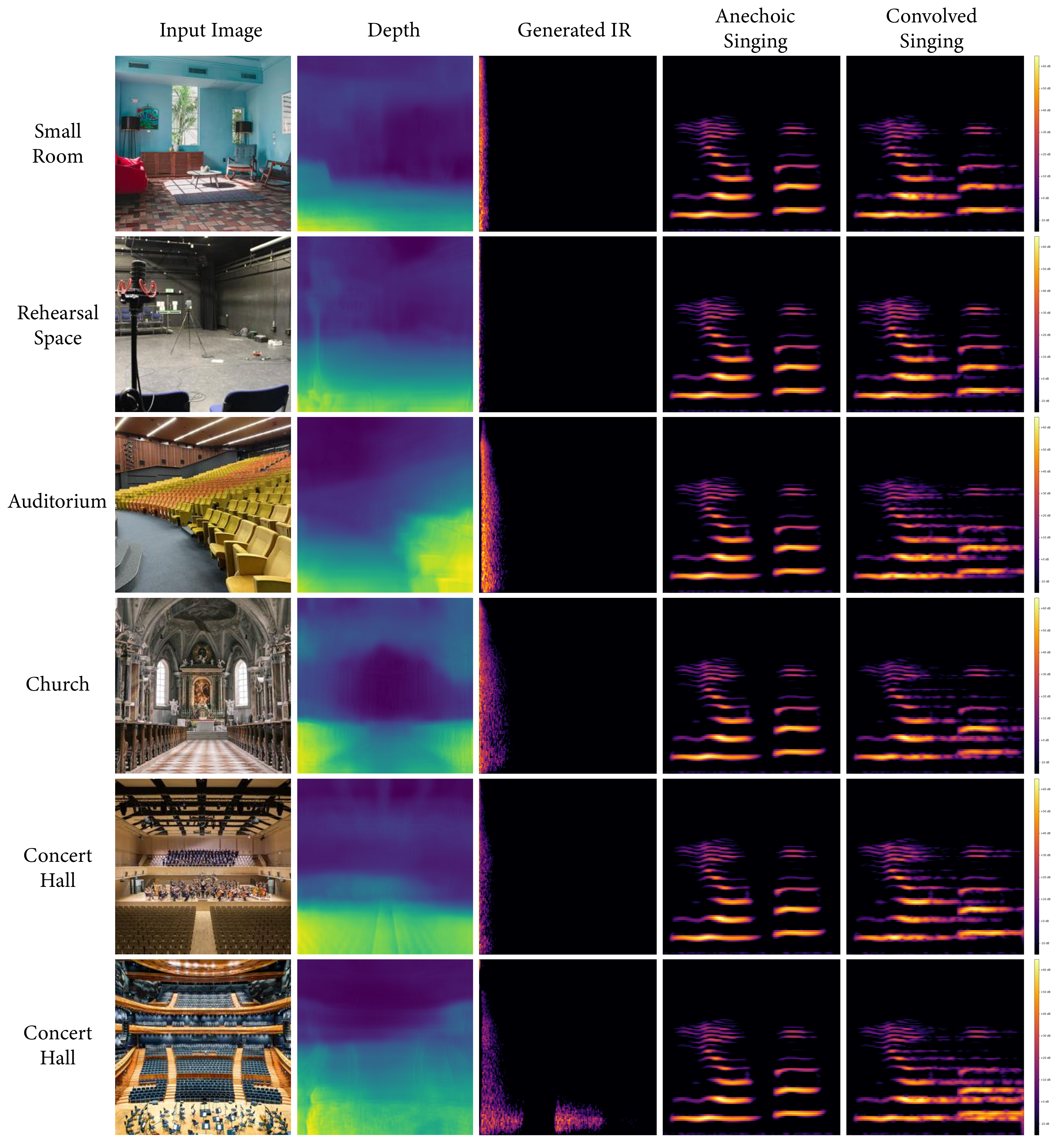}
    \caption{Music. Columns show input images, depth maps, generated IRs, and an anechoic vocal singing signal before and after the generated IR was applied to the signal via convolution. The input images come from spaces relevant to music including a typical small room, an acoustically treated rehearsal space, an auditorium, a church, and 2 large concert halls. Generally, larger spaces tend to exhibit longer decay times in the output, however some examples such as the concert halls with visible acoustic treatment appear to have a shorter decay than more reverberant spaces like the church or auditorium with more reflective surfaces. The final concert hall shows an atypical impulse response with a visible discontinuity in the IR tail. This is not commonly observed among our model outputs, but illustrates the nature of artifacts which can occasionally occur.}
    \label{fig:p_music}
\end{figure*}

\begin{figure*}
    \centering
    \includegraphics[width=0.82\textwidth]{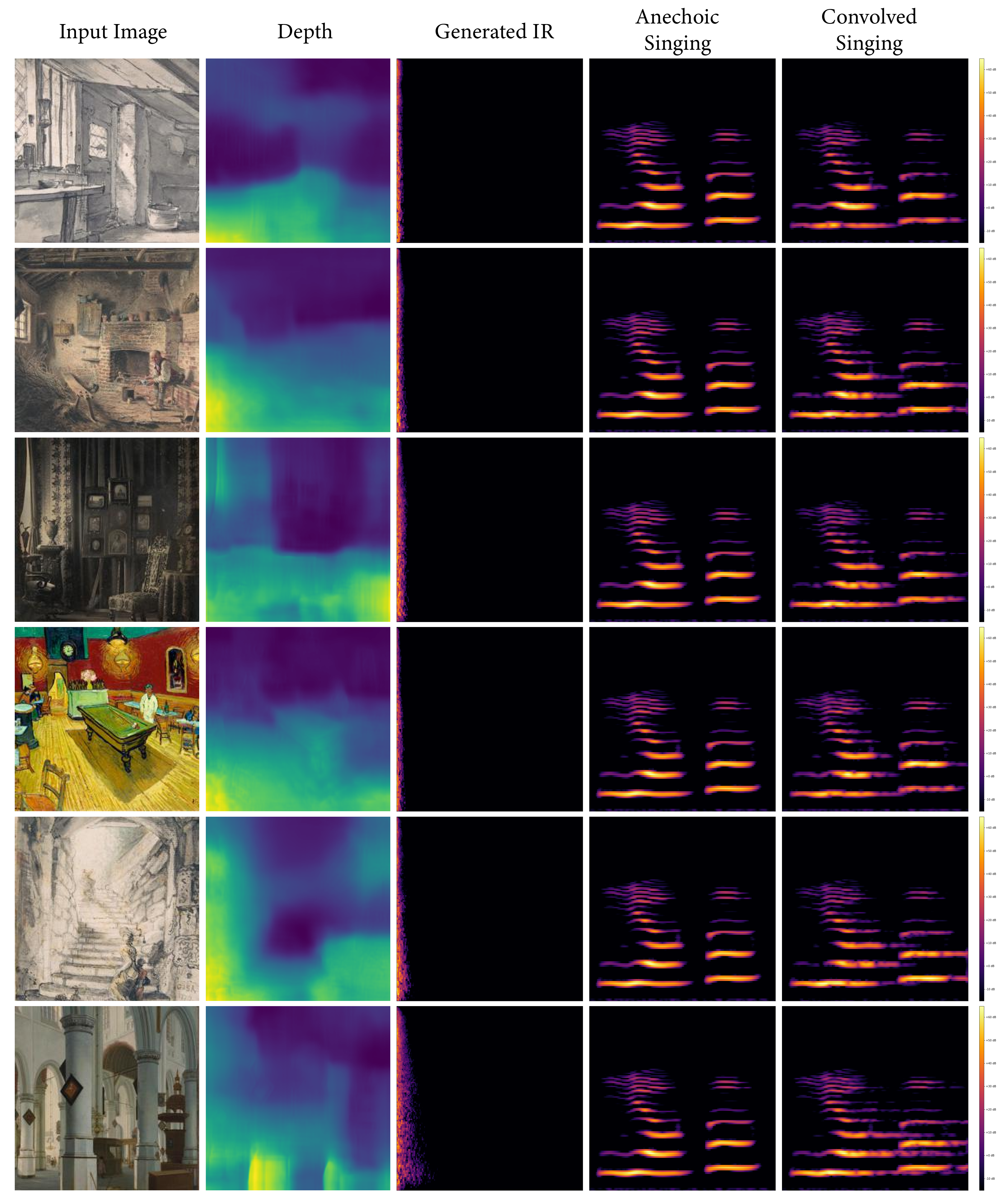}
    \caption{Art. Columns show input images, depth maps, generated IRs, and an anechoic operatic singing signal before and after the generated IR was applied to the signal via convolution. Images here are drawings, paintings and a vintage art photograph ca. 1850. Artistic depictions of spaces were not included in our training dataset. In many cases, plausible impulse responses are generated from such input images. In general, larger depicted spaces, like the church in the bottom row, exhibit longer decay times as is observed with standard 2D photographs.}
    \label{fig:p_art}
\end{figure*}

\begin{figure*}
    \centering
    \includegraphics[width=\textwidth]{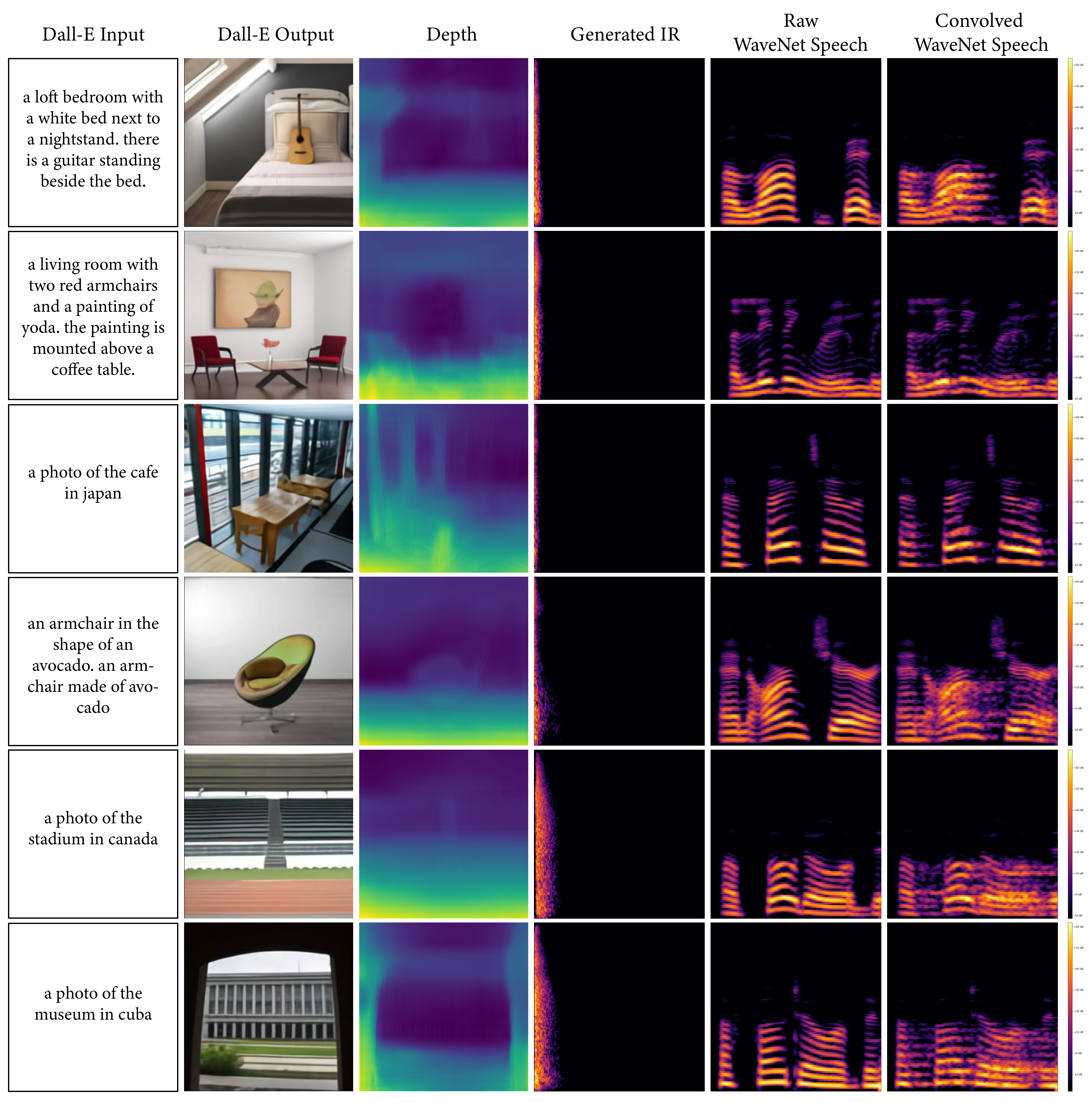}
    \caption{DALL·E. Images generated from text by DALL·E \cite{dalle} used here as input images. The same corresponding input text was synthesized via text-to-speech as our signal of interest and convolved with the generated IR.  This reflects synthetic speech in a synthetic environment, indicating a path for synthesizing realistic IRs from text. It also shows how our model might work with other state-of-the-art generative media models to produce more consistent and realistic results in different domains.}
    \label{fig:p_dall-e}
\end{figure*}

\begin{figure*}
    \centering
    \includegraphics[width=0.85\textwidth]{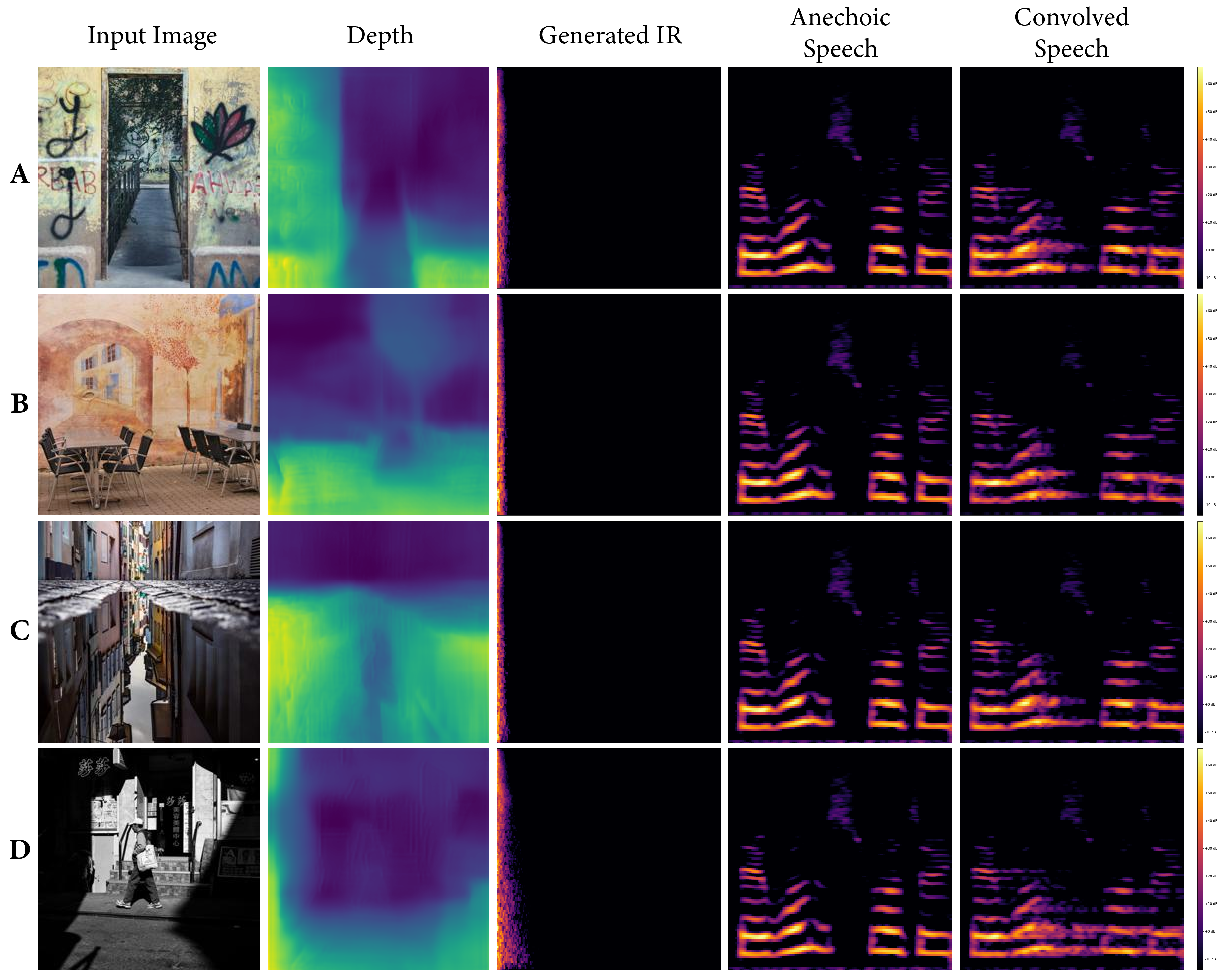}
    \caption{Challenging images. Input images containing street murals, reflections, and shadows demonstrating cases where depth is inaccurately estimated. (A) A painted doorway giving the illusion of depth. (B) A wall with a mural of a street and tree where the depth of the wall is inaccurately estimated. (C) A low-angle photo of a reflective puddle. (D) An outdoor street image with strong shadows which results in a depth map and generated IR more similar to a room than an outdoor space. These more extreme scenarios are chosen to clearly illustrate the limitations of our approach.}
    \label{fig:p_reflection}
\end{figure*}

\begin{figure*}
    \centering
    \includegraphics[width=0.85\textwidth]{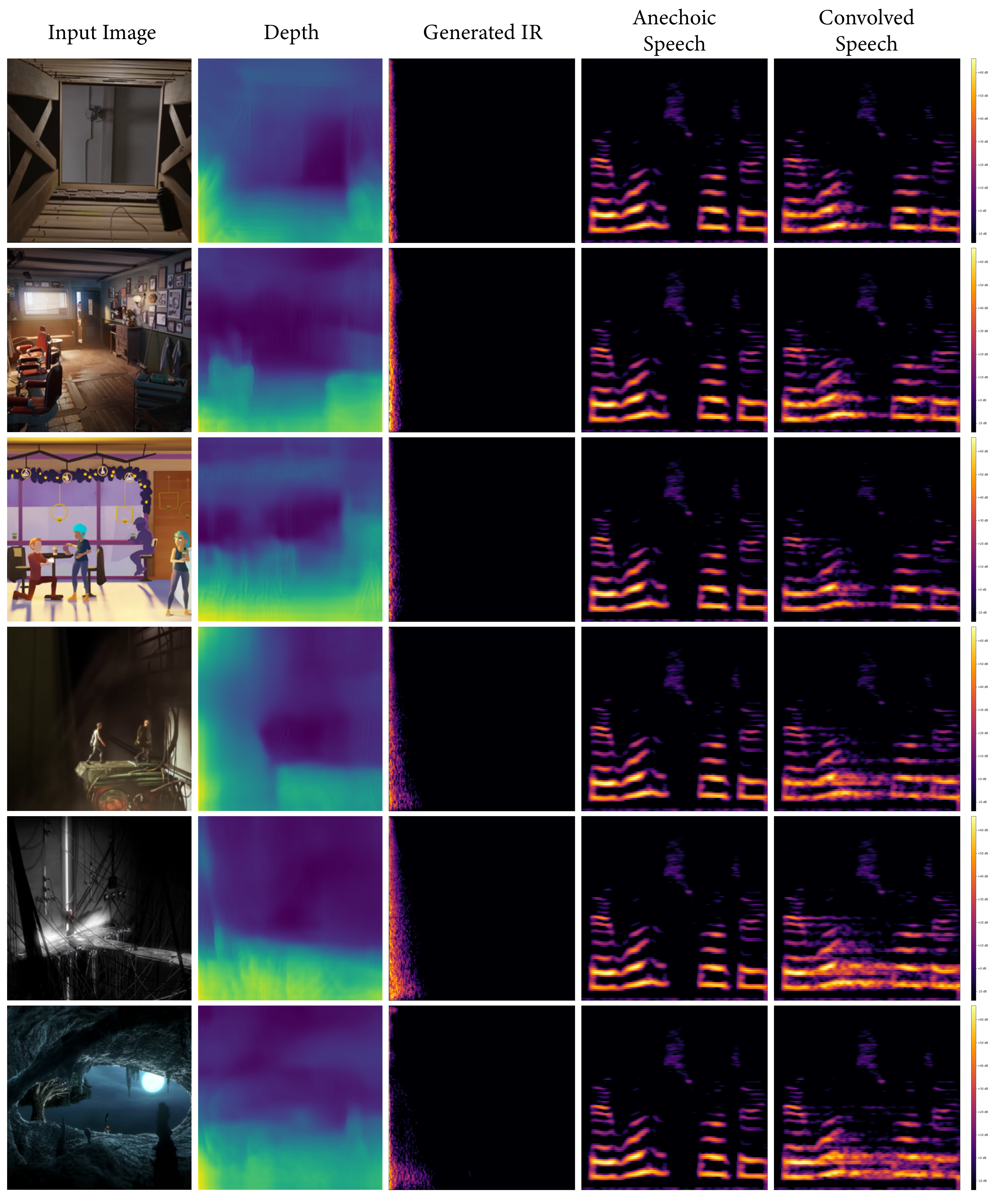}
    \caption{Animated films. Scenes from Blender open animation films used as input images (speech convolved with generated IRs). Columns show input image, calculated depth map, spectrogram of generated IR, an anechoic passage reading sample, and the same passage with the generated IR applied via convolution. In general, we find that our model plausibly estimates the reverberant characteristics of these spaces. For example, the wooden small space is very brief. The barbershop appears longer due to some artefacts, but the broadband decay is relatively quick as can be heard in the audio. Seemingly larger spaces again correspond to longer IRs. This is a case of Real2Sim transfer, where we can approximate IRs directly that sound as measured IRs, but in virtual environments where this measurement is not possible.}
    \label{fig:p_animation}
\end{figure*}

\begin{figure*}
    \centering
    \includegraphics[width=0.85\textwidth]{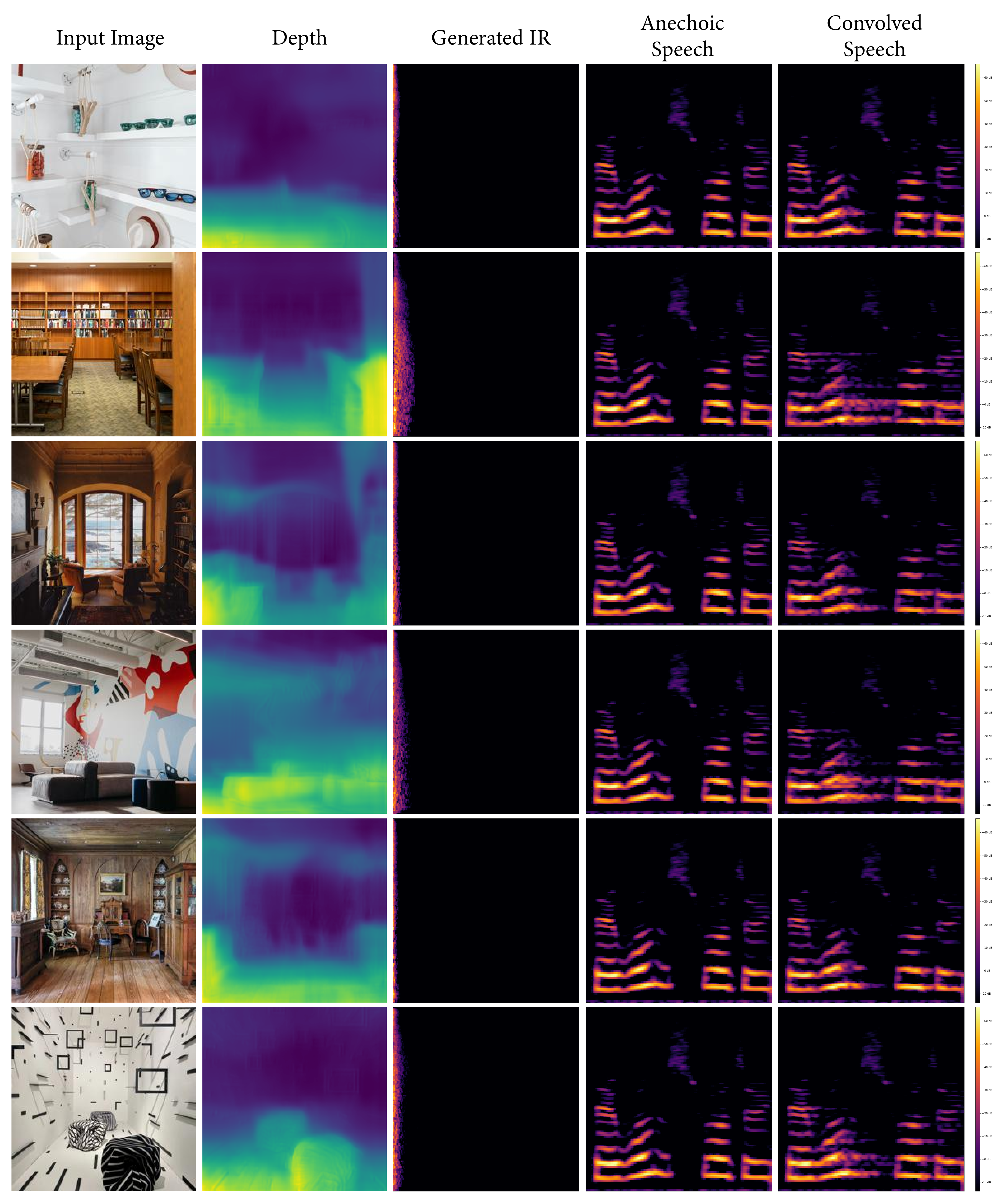}
    \caption{Virtual backgrounds. Images which may serve as virtual backgrounds used as input images to our model. These reflect spaces that may be used for videoconferencing or other online meetings. Realistic IRs may be generated and used in these contexts to increase the sense of being in a shared space with others.}
    \label{fig:p_zoom}
\end{figure*}

\begin{figure*}
    \centering
    \includegraphics[width=0.9\textwidth]{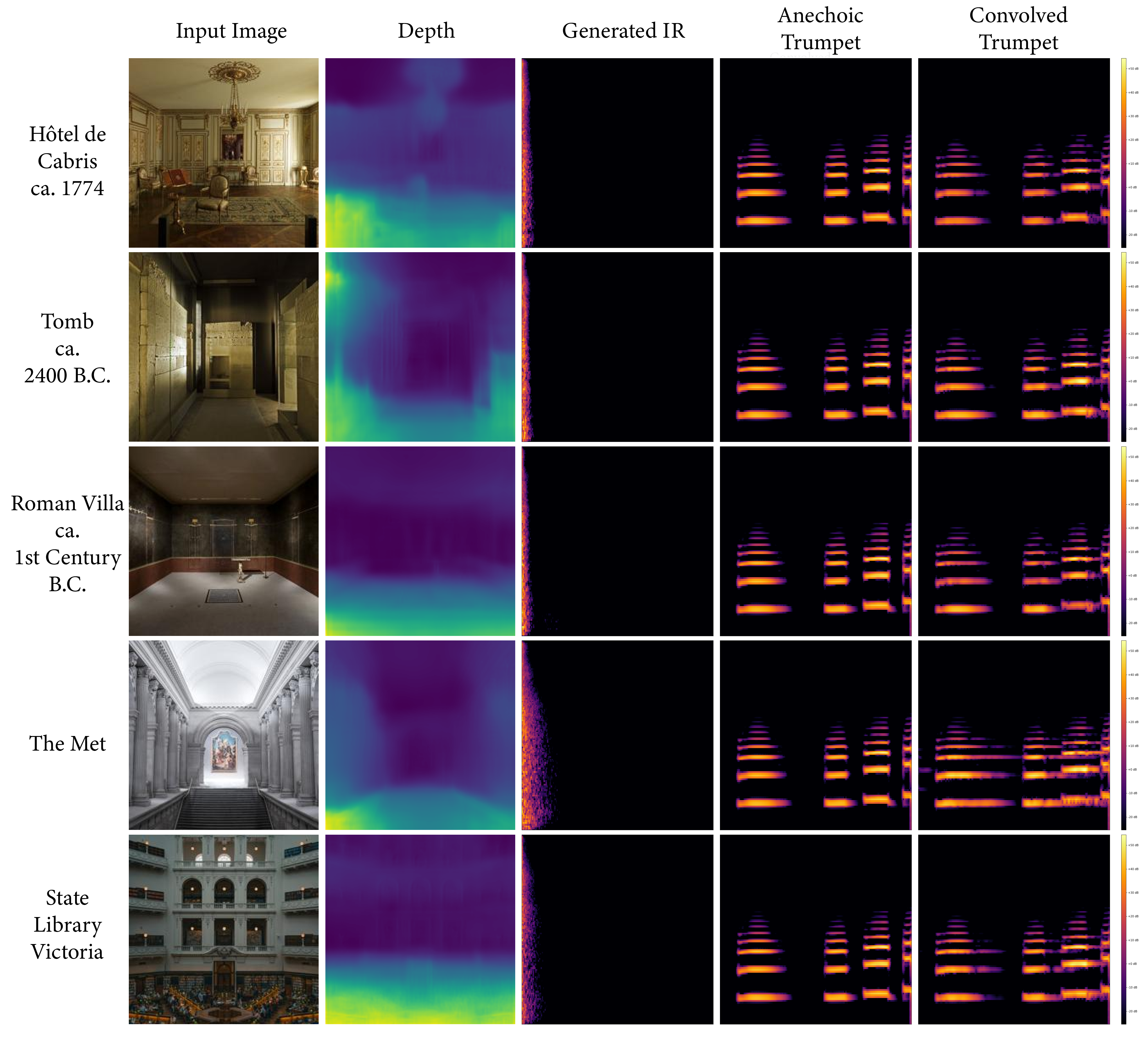}
    \caption{Historical and notable places. Additional examples of unusual and historical spaces which may be difficult or impossible to obtain IRs from.}
    \label{fig:p_other}
\end{figure*}

\begin{figure*}
    \centering
    \includegraphics[width=0.9\textwidth]{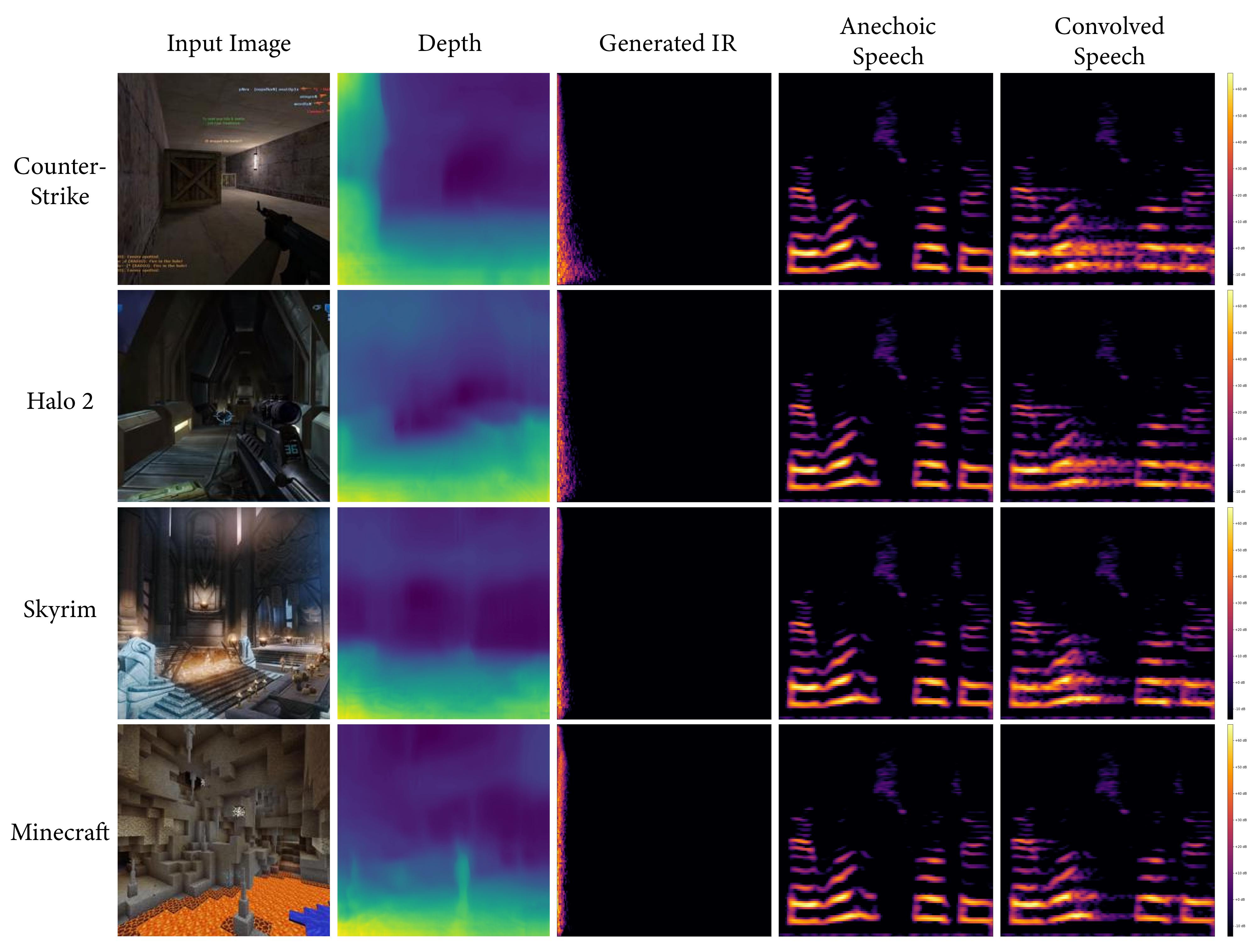}
    \caption{Video games. Impulse responses generated and applied via convolution from screenshots of four 3D video games. Video games are one example of a virtual space that might benefit from easily generated impulse responses. While the medium sized room from Counter-Strike and the large hallway from Halo 2 may be plausible IRs, the large hall shown in the Skyrim screenshot and the cavern in the Minecraft example do not have correspondingly long reverberant tails as would be expected showing possible examples of where the scale of the space was not accurately estimated. 3D rendered images were not included in our dataset but are a ripe area of future work which might greatly increase the performance of our model on both real scenes and virtual scenes such as these video game examples.}
    \label{fig:p_videogame}
\end{figure*}

\begin{figure*}
    \centering
    \includegraphics[width=0.9\textwidth]{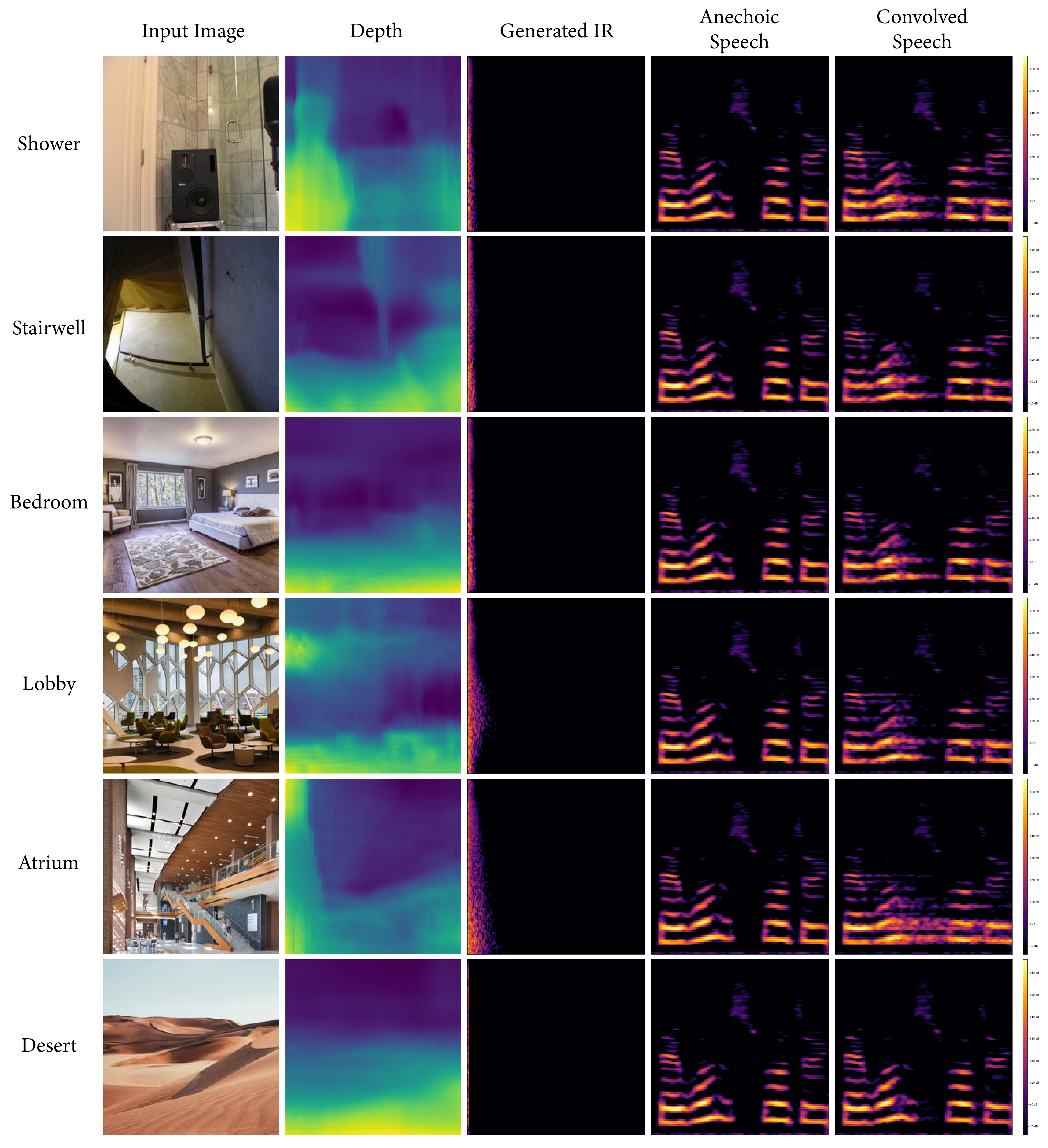}
    \caption{Common and identifiable scenes. Input images and the resulting IRs are shown and convolved with an anechoic speech signal. Input images here reflect spaces that are regularly encountered in everyday life yet may not often be recorded in. These types of scenes are useful for audio post-production as they may be commonly found in movies and television shows. Small and outdoor scenes are observed to have very brief IRs while in comparison, the larger building interior has a much longer output IR as expected.}
    \label{fig:p_everyday}
\end{figure*}

